\documentclass{pnastwo}

\usepackage{graphicx}

\usepackage{amssymb,amsfonts,amsmath}

\contributor{ }
\url{ }
\copyrightyear{ }
\issuedate{ }
\volume{ }
\issuenumber{ }

\begin{document}



\title{From antiferromagnetic insulator to correlated metal in pressurized and doped LaMnPO}




\author{J. W. Simonson\affil{1}{Department of Physics and Astronomy, Stony Brook University, Stony Brook, NY 11794, USA}, Z. P. Yin\affil{1},\affil{2}{Department of Physics and Astronomy, Rutgers University, Piscataway, NJ 08854, USA}, M. Pezzoli\affil{1},\affil{2}, J. Guo\affil{3}{Institute of Physics and Beijing National Laboratory for Condensed Matter Physics, Chinese Academy of Sciences, Beijing 100190, P. R. China}, J. Liu\affil{4}{Institute of High Energy Physics, Chinese Academy of Sciences, Beijing 100039, P. R. China}, K. Post\affil{5}{Department of Physics, University of California, San Diego, La Jolla, CA 92093-0319, USA}, A. Efimenko\affil{6}{Max-Planck-Institut f\"{u}r Chemische Physik fester Stoffe, D-01187 Dresden, Germany}, N. Hollmann\affil{6}, Z. Hu\affil{6}, H.-J. Lin\affil{7}{National Synchrotron Radiation Research Center (NSRRC), 101 Hsin-Ann Road, Hsinchu 30077, Taiwan}, C. T. Chen\affil{7}, C. Marques\affil{1}, V. Leyva\affil{1}, G. Smith\affil{1}, J. W. Lynn\affil{8}{NIST Center for Neutron Research, Gaithersburg, MD 20899, USA}, L. Sun\affil{3}, G. Kotliar\affil{2}, D. N. Basov\affil{5}, L. H. Tjeng\affil{6}, \and M. C. Aronson\affil{1},\affil{9}{Brookhaven National Laboratory, Upton, NY 11973, USA}}
\contributor{ }

\maketitle

\begin{article}

\begin{abstract}Widespread adoption of superconducting technologies requires the discovery of new materials with enhanced properties, especially higher superconducting transition temperatures T$_{c}$. The unexpected discovery of high T$_{c}$ superconductivity in cuprates and in materials as diverse as heavy fermions, organic conductors, and endohedrally-doped fullerenes suggests that the highest T$_{c}$s occur when pressure or doping transform the localized and moment-bearing electrons in antiferromagnetic insulators into itinerant and weakly magnetic metals. The absence of this delocalization transition in Fe-based superconductors may limit their T$_{c}$s, but even larger T$_{c}$s  may be possible in their isostructural Mn analogs, which are antiferromagnetic insulators like the cuprates. It is generally believed that prohibitively large pressures would be required to suppress the strong Hund's rule coupling in these Mn-based compounds, collapsing the insulating gap and enabling superconductivity. Indeed, no Mn-based compounds are known to be superconductors. The electronic structure calculations and x-ray diffraction measurements presented here challenge these long held beliefs, finding that only modest pressures are required to transform LaMnPO, isostructural to superconducting host LaFeAsO, from an insulating tetragonal structure with a large Mn moment to a gapless orthorhombic structure with a small Mn moment. Proximity to this electronic delocalization transition in LaMnPO results in a highly interacting metallic state, the familiar breeding ground of superconductivity.

\end{abstract}

\keywords{correlated electron systems | superconductivity | electronic delocalization transition}





\dropcap{S}uperconductivity with high transition temperatures T$_{c}$ was first found near an electron delocalization transition (EDT) in the cuprates, and subsequently in compounds as diverse as quasi-two dimensional organic layers~\cite{kanoda1997}, heavy fermions~\cite{shishido2005,sarrao2007}, and endohedrally doped fullerides~\cite{takabayashi2009}.  One obstacle to achieving a higher T$_{c}$ in the Fe-based superconductors may be that the parent compounds are metallic~\cite{ishida2009,johnston2010,paglione2010}, albeit with quasiparticle mass enhancements~\cite{yin2011a} that suggest varying degrees of proximity to an EDT~\cite{qazilbash2009,si2008,basov2011}.  So far no insulating parent compounds have been identified that can, by analogy to the cuprates, be doped to achieve higher superconducting transition temperatures, although is possible that the recently isolated K$_{2}$Fe$_{4}$Se$_{5}$~\cite{song2011} and La$_{2}$O$_{2}$Fe$_{2}$O(Se,S)$_{2}$~\cite{zhu2010} phases may prove to be the first compounds of this type. In contrast, isostructural Mn-based compounds often have large insulating gaps and ordered moments~\cite{bronger1984,continenza2001}, suggesting their suitability as possible parent compounds. However, at present there are no known Mn-based superconductors,  and it is generally believed that the Hund's rule coupling in Mn compounds is prohibitively strong, so that doping will not reduce the correlations to the point at which superconductivity may become possible. The electronic structure calculations and x-ray diffraction measurements presented here find that only modest pressures are required to transform LaMnPO, isostructural to the superconducting host LaFeAsO, from an insulating tetragonal structure with a large Mn moment to a gapless orthorhombic structure with a small Mn moment.  Our theoretical and experimental investigations of \mbox{LaMnPO} show that Mn-based correlation gap compounds can be surprisingly close to electronic delocalization, and the result may well be the stabilization of the same strongly correlated metallic state that is thought to be an essential ingredient for unconventional superconductivity in many other classes of materials.

We combine first-principles electronic structure calculations with spectroscopic and diffraction measurements to show that high pressures but not electron doping drive an EDT in single crystals of the magnetic insulator \mbox{LaMnPO}. We have selected \mbox{LaMnPO} for this study because it forms in the same ZrCuSiAs structure as the superconducting parent compound LaFeAsO, consisting of functional Mn$^{2+}$P$^{3-}$ layers stacked with charge donor La$^{3+}$O$^{2-}$ layers~\cite{nientiedt1997}. Electronic structure calculations using the generalized-gradient approximation with varying Hubbard U (GGA+U) within density functional theory (DFT) confirmed the insulating gap revealed in initial  electrical resistivity, optical conductivity, and photoemission measurements performed on polycrystalline \mbox{LaMnPO} ~\cite{yanagi2009,yanagi2010}.  However, GGA+U calculations have had limited success in reproducing experimental observations, and significant improvement is found when a combination of DFT and dynamical mean field theory (DFT+DMFT) is used~\cite{yin2011a,yin2011b}. We show here that the insulating and magnetic character of \mbox{LaMnPO} is well captured by DFT+DMFT calculations~\cite{kotliar2006} by making direct contact to a variety of measurements performed on high quality single crystals. With these improvements, a new picture of insulating \mbox{LaMnPO} emerges, where substantial charge fluctuations suggest a nearby EDT.  No EDT is found in electron-doped \mbox{LaMnPO}$_{1-x}$F$_{x}$, but high pressure x-ray diffraction measurements show that \mbox{LaMnPO} undergoes a structural transition and a subsequent volume collapse at pressures where electronic structure calculations find that the Mn moments are themselves close to collapse in \mbox{LaMnPO}.

\begin{figure}
\noindent\includegraphics[width=8.7cm]{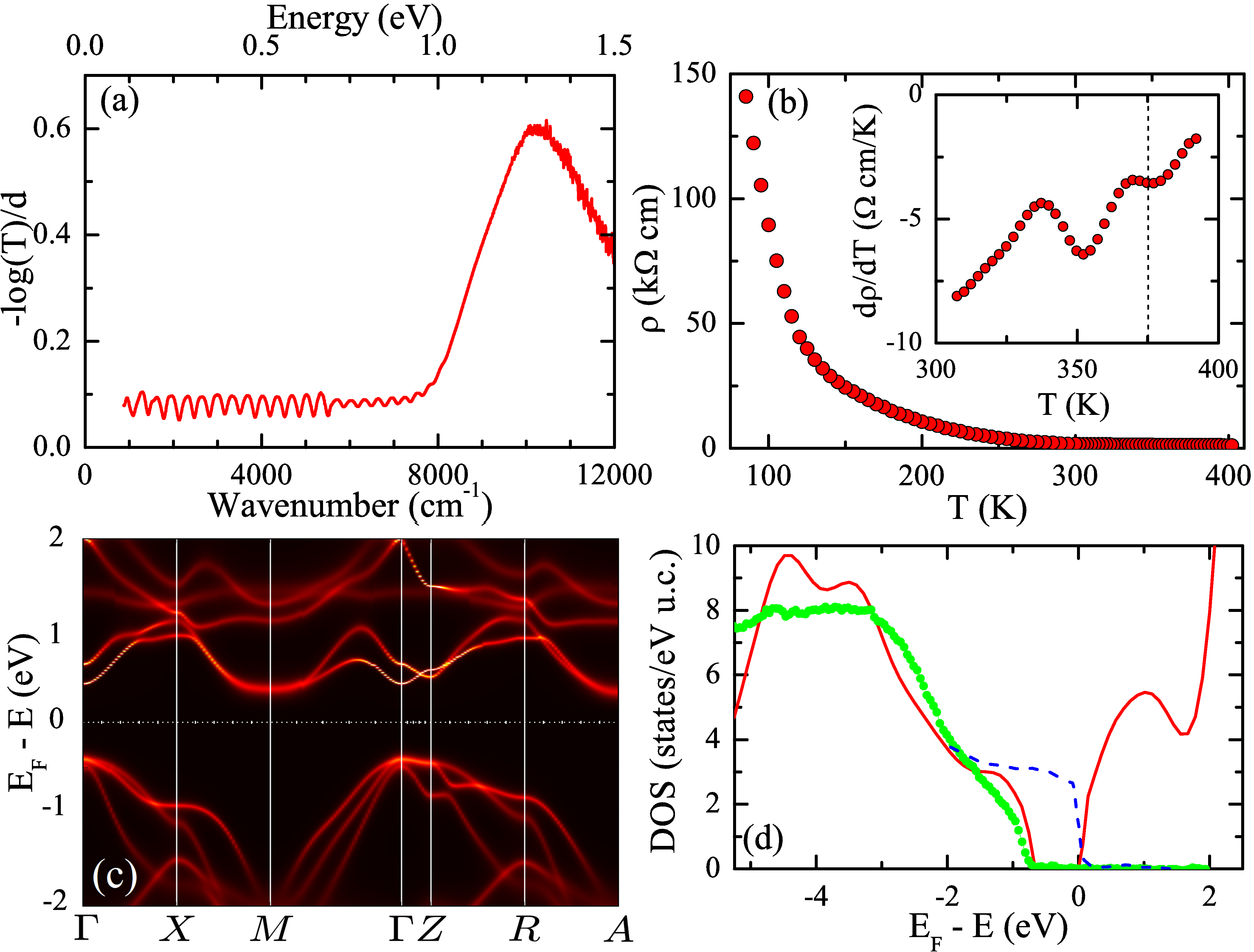}
\caption{Correlation gap in single crystal \mbox{LaMnPO}. a, The wavenumber (bottom axis) and energy (top axis) dependencies of the optical transmission T($\omega$) measured at 300 K. The sample thickness d=6.9 microns. b, Temperature dependence of the electrical resistivity $\rho$(T) in single crystal \mbox{LaMnPO}. Inset: expanded view of d$\rho$/dT shows a weak ordering anomaly near T$_{N}$=375 \emph{K} (vertical dashed line). c, The theoretical photoemission spectrum of \mbox{LaMnPO} from DFT+DMFT. d, A comparison of the energy dependencies of the total densities of states determined from DFT+DMFT calculations (red line) that have been broadened to represent the energy resolution of angle integrated photoemission measurements, carried out here on single crystals (green points). Measurements of Ag (blue dashed line) place the Fermi level E$_{F}$ near the bottom of the calculated conduction band. \label{insulator}}
\end{figure}

The insulating character of \mbox{LaMnPO} is evident from the optical transmission T($\omega$) (Fig.~1a), where a rapid increase near 8000 cm$^{-1}$ reveals the optical gap $\Delta\simeq$1.3 eV.  The electrical resistivity $\rho$(T) is decidedly insulating, rising from  $\approx$1200 $\Omega$-cm at 400 K to $\approx$140,000 $\Omega$-cm at 85 K (Fig.~1b)~\cite{simonson2011}.  The theoretical photoemission spectrum (Fig.~1c) shows that there is a direct gap of 0.8 eV at $\Gamma$, and a smaller indirect gap of 0.65 eV along the $\Gamma$-A direction. The total density of states determined from these calculations compares favorably to angle-integrated photoemission experiments (Fig.~1d), finding that the top of the valence band is 0.9 eV below the Fermi level, which is located at the bottom of the conduction band. Overall, DFT+DMFT provides an accurate account of the electronic gaps obtained from photoemission and optical measurements.

\begin{figure}
\noindent\includegraphics[width=8.7cm]{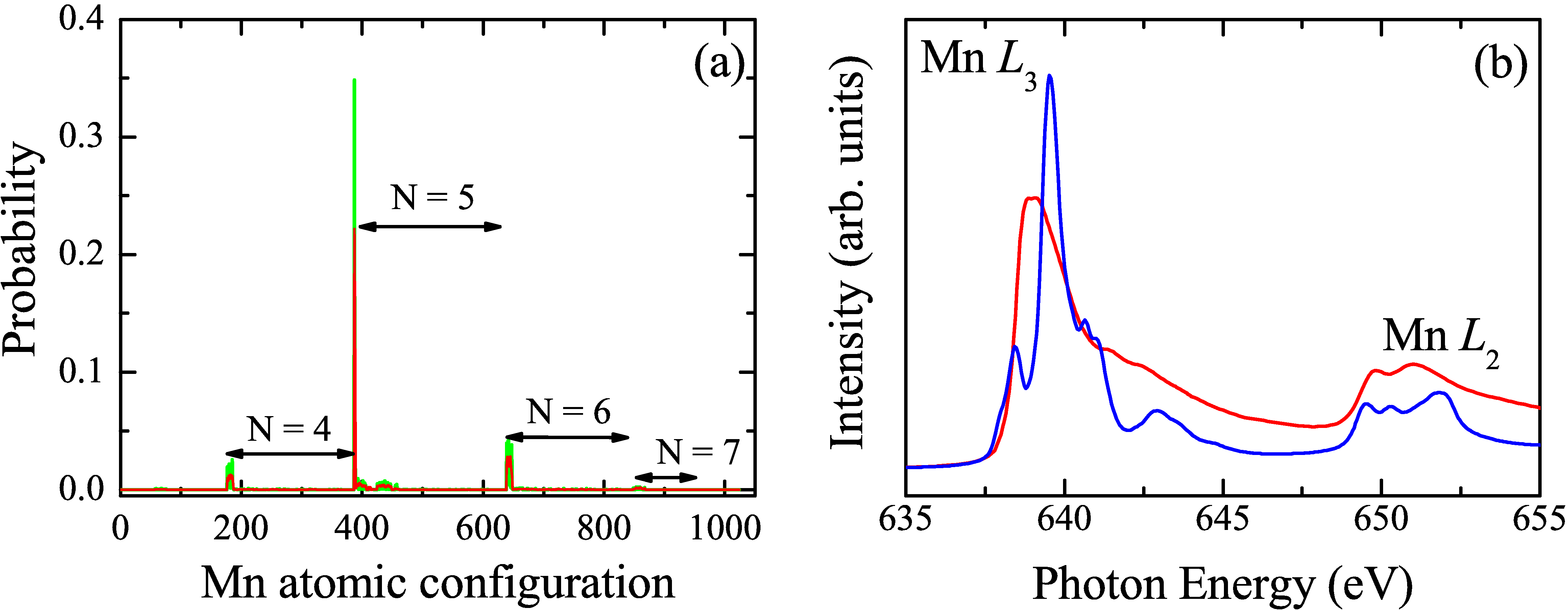}
\caption{Charge Fluctuations in \mbox{LaMnPO}. a, The histogram of atomic states comprising the Mn-3d shell for \mbox{LaMnPO}, deduced from the DFT+DMFT calculations performed in the paramagnetic (red) and magnetically ordered (green) states. There are a total of 1024 states for Mn atoms with up to 10 electrons in the 3d shell, each with different energy and occupancy N. b, The Mn L$_{2;3}$  x-ray absorption spectra of LaMnPO (red) and MnO (blue) obtained at 300 \emph{K}. \label{valence}}
\end{figure}

It is generally believed that valence fluctuations are negligible in Mn$^{2+}$ compounds such as LaMnPO, since a strong Hund's rule interaction aligns the spins of all five d-electrons, yielding a half-filled d-shell with a net Mn moment of 5$\mu_{B}$. The atomic histogram of the Mn 3d shell (Fig.~2a) shows the probability for the Mn atom to be in each of its 1024 states of different energy and occupancy. As expected, the N=5 states are the most highly occupied, but there is considerable weight in the states with N$\neq$5, suggesting that there are substantial valence fluctuations in \mbox{LaMnPO} that are intermediate in strength between those of the iron pnictides~\cite{yin2011a} and the stable valence cuprates~\cite{shim2007}. X-ray absorption measurements at the Mn $L_{2,3}$ edges support these findings. Fig.~2b displays the spectrum of \mbox{LaMnPO}
together with that of MnO, which is taken as a reference material with the ionic $3d^{5}$ Mn configuration. One can clearly observe that the main peak of the \mbox{LaMnPO} L$_{3}$ white line is at a lower energy that that of MnO, and that the multiplet-derived features are broader, indicating the presence of appreciable charge fluctuations in LaMnPO.

The antiferromagnetic order parameter (Fig.~3a) establishes that the Ne\'{e}l temperature T$_{N}$ =375 \emph{K} $\pm$ 5 \emph{K} and that the ordered moment is 3.28$\pm$0.05 $\mu_{B}$/Mn for T$\rightarrow$0, in excellent agreement with the ordered moment of 3.2 $\mu_{B}$/Mn that is calculated from DMFT+DFT. In the paramagnetic state, these calculations find a moment magnitude of $\sim$4 $\mu_{B}$/Mn, and this reduction from the high spin state value of 5 $\mu_{B}$/Mn is further evidence for the importance of charge fluctuations. Interestingly, there is no indication of a Curie-Weiss contribution to the magnetic susceptibility $\chi$ for T$\geq$T$_{N}$, since $\chi$ is nearly temperature independent for temperatures above and below T$_{N}$ (Fig.~3b). This absence of individual, fluctuating Mn moments in the paramagnetic state suggests that strong exchange coupling J dynamically compensates the Mn moments, and within the context of the t-J model~\cite{johnston1989,singh1992,auerbach1988}, the weak temperature dependence implies that T/J$\ll$1, even when T=800 K. In particular, no feature was observed in $\chi$(T) at the 375 K ordering temperature (inset, Fig.~3b), although it was measured on the same sample used for neutron diffraction. A weak upturn in $\chi$ is found at the lowest temperatures, and the inset of Fig.~3c shows that it can be fit for T$\leq$50 \emph{K} to the Curie-Weiss expression $\chi$(T) = C/(T-$\theta$) with $\mu_{CW}$=0.34$\pm$0.03 $\mu_{B}$/Mn and Weiss temperature $\theta$= -5$\pm$2 \emph{K}.

\begin{figure}
\noindent\includegraphics[width=8.7cm]{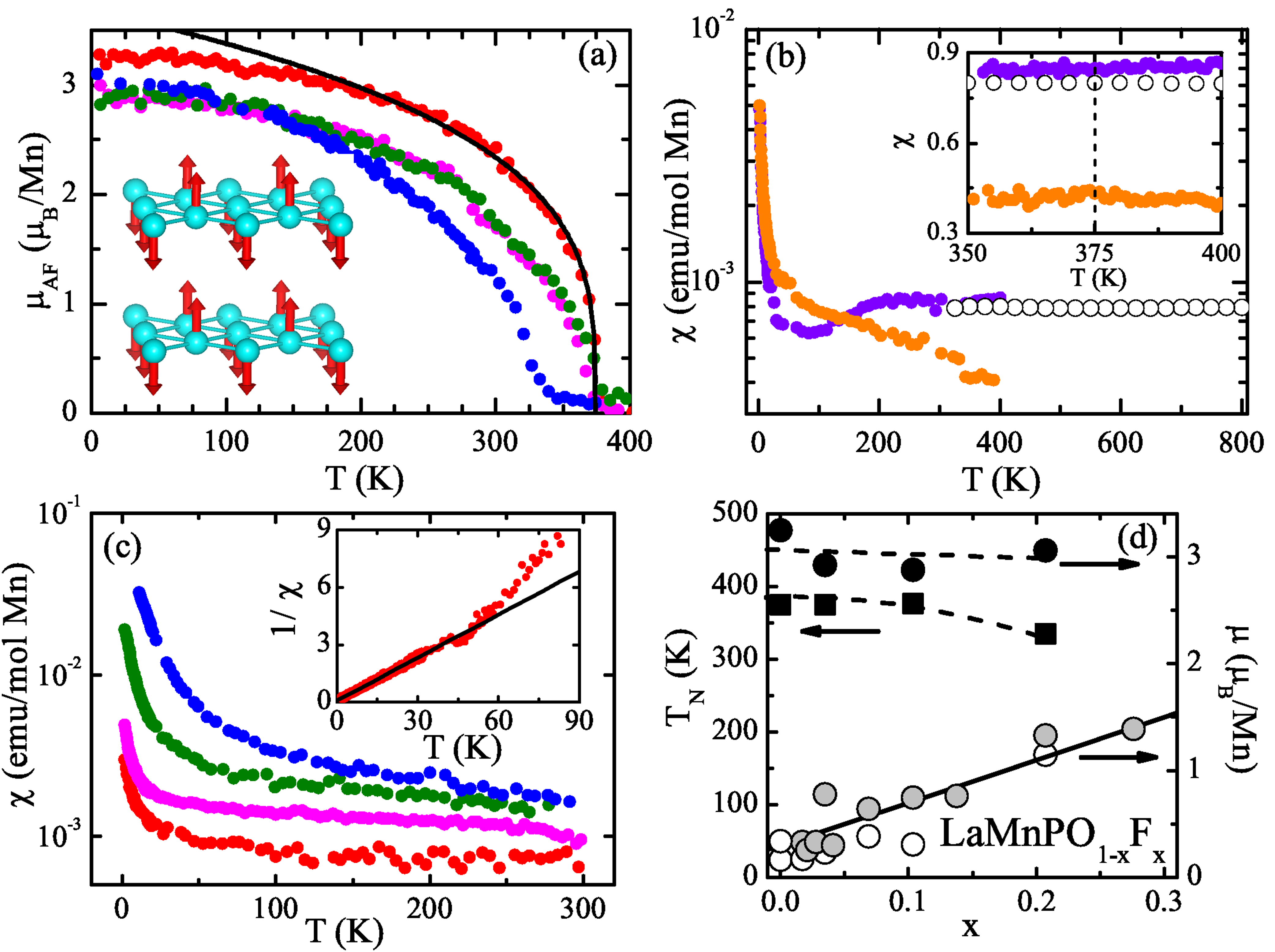}
\caption{Magnetic Moments in \mbox{LaMnPO}. a, Temperature dependencies of the ordered Mn moment $\mu_{AF}$ in LaMnP(O$_{1-x}$F$_{x}$) for x=0 (red), 0.07 (magenta), 0.1 (green), and 0.21 (blue), taken from the temperature dependence of the (100) magnetic peak in neutron diffraction measurements. Solid line is a fit to the power law M$\simeq$(T$_{N}$-T)$^{-\beta}$, with $\beta$=0.28$\pm$0.02, in reasonable agreement with the Ising ($\beta$=0.326) or Heisenberg ($\beta$=0.365) exponents~\cite{zinnjustin1977} Inset: the magnetic structure of \mbox{LaMnPO} consists of \emph{ab} planes where the Mn moments are arranged in a checkerboard pattern. These planes are stacked ferromagnetically along the \emph{c-axis}, as previously reported~\cite{yanagi2009}. b, The dc magnetic susceptibility $\chi$(T)=M/B (B=1 T) for B$\parallel$c (purple),B$\perp$c (orange), and polycrystalline \mbox{LaMnPO} (white). Inset: expanded view near T$_{N}$, indicated by vertical dashed line. $\chi$(T) shows only weak anisotropy, and no ordering anomaly is observed near T$_{N}$ for B$\parallel$c, B$\perp$c, or for the random orientation of the polycrystalline sample. c, Temperature dependencies of $\chi$(T) single crystals of LaMnPO$_{1-x}$F$_{x}$, with different values of x as in (a).  Inset: temperature dependence of the inverse susceptibility 1/$\chi$ of \mbox{LaMnPO}. Solid line is a fit to Curie-Weiss expression for T$\leq$50 \emph{K}. d, Variation of T$_{N}$(filled squares), $\mu_{AF}$(filled circles), and $\mu_{CW}$ (open circles: polycrystalline LaMnPO, gray circles: single crystal LaMnPO) with measured F concentration x, where the slope of the solid black line shows that the fluctuating moment $\mu_{CW}$increases by $\simeq$3.8$\pm$0.3 $\mu_{B}$ per electron added. \label{magnetic}}
\end{figure}

The data in Fig.~3a demonstrate that even heavy electron doping in \mbox{LaMnPO}$_{1-x}$F$_{x}$  has  little effect on either T$_{N}$ or $\mu_{AF}$ (Fig.~3d).  Measurements of $\chi$(T) (Fig.~3c) reveal, however, that the Curie-Weiss tail that was observed in pure \mbox{LaMnPO} below $\simeq$50 \emph{K} grows dramatically with doping, where the corresponding fluctuating moment $\mu_{CW}$ increases at the rate of $\simeq$3.8$\pm$0.3 $\mu_{B}$ per electron added (Fig.~3d), in close agreement with the 4.0 $\mu_{B}$/Mn moment predicted by DFT+DMFT in the absence of magnetic order. The emerging picture is that electron doping breaks the strong exchange coupling among Mn moments that is responsible for the temperature independence of  $\chi$(T), and the absence of additional magnetic ordering in neutron diffraction measurements implies that these newly single moments continue to fluctuate freely at temperatures as low as 4 \emph{K}.

Taken together, these results indicate that \mbox{LaMnPO} may not be as electronically stable as its substantial gap and ordered moment suggest, and indeed may be close to an EDT that is driven by the nucleation of states with energies within the correlation gap, and not by the collapse of the gap itself~\cite{kotliar2006}. Within this description of the EDT, the in-gap states are initially localized and moment-bearing, but as the system is tuned towards the EDT by pressure or doping, they eventually delocalize to form a metal that becomes progressively less correlated and less magnetic away from the EDT.  In accordance, optical transmission measurements find that even substantial  electron doping in \mbox{LaMnPO}$_{1-x}$F$_{x}$ does not appreciably reduce the correlation gap~\cite{simonson2011}, while neutron diffraction measurements show that the ordered Mn moment is similarly robust. Electrical resistivity and optical conductivity measurements document the buildup with electron doping of in-gap states with localized charge~\cite{simonson2011}, while magnetic susceptibility measurements find that each doped electron introduces an uncoupled and fluctuating moment $\mu_{CW}\simeq$ 3.8 $\mu_{B}$, which is close to the full Mn moment expected from DFT+DMFT calculations. We propose that these in-gap states are the localized precursors of a potentially metallic state that would form if the Hund's rule correlations were weakened enough to allow these moments and charges to hybridize into strongly correlated bands, as may occur in SmMnAsO$_{1-x}$\cite{shiomi2011}, but not in BaMn$_{2}$As$_{2}$ or PrMnSbO~\cite{johnston2010}. Apparently \mbox{LaMnPO} is simply too far from an EDT to be driven metallic by even the large amounts of electron doping reported here.

\begin{figure}
\noindent\includegraphics[width=8.7cm]{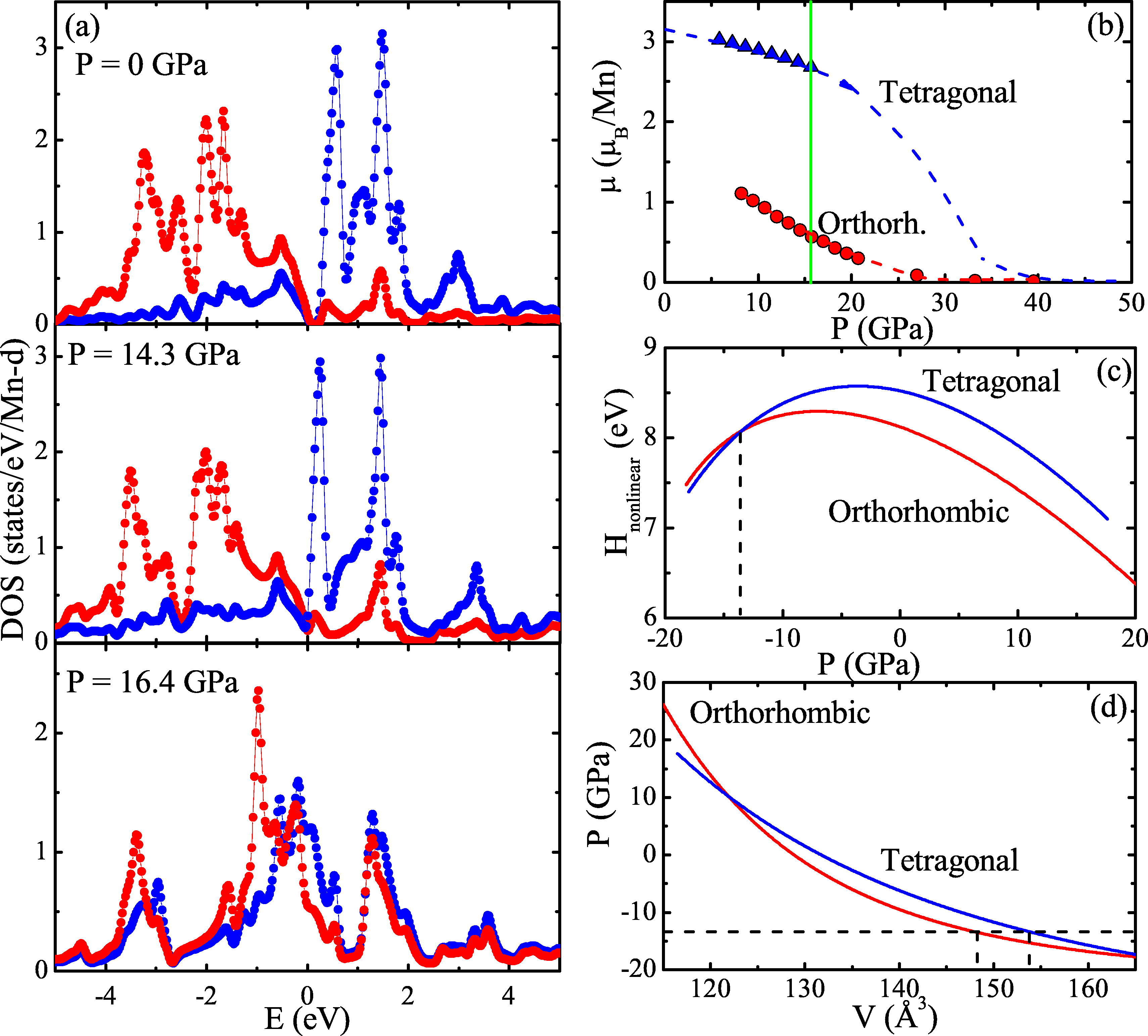}
\caption{LSDA calculations of \mbox{LaMnPO} under pressure.  a, Density of states plots for tetragonal \mbox{LaMnPO} at 0 \emph{GPa} (top) and 14.3 \emph{GPa} (middle) and orthorhombic \mbox{LaMnPO} at 16.4 \emph{GPa} (bottom) from LSDA calculations carried out at the experimental volumes that correspond to the indicated pressures. Contributions from up spin states are shown in red and down spin states in blue.  b, The antiferromagnetic moment $\mu$ as a function of pressure from LSDA calculations for \mbox{LaMnPO} in the tetragonal (blue) and orthorhombic (red) structures. The tetragonal-orthorhombic transition is experimentally found at 16 \emph{GPa} (green line). Blue and red dashed lines through data are guides for the eye.  c, The computed nonlinear part of the enthalpy, H$_{nonlinear}$=H - (P/31.98 \emph{GPa} - 543.5 keV) as a function
of pressure for the tetragonal and orthorhombic phases. They cross at the critical pressure P$_{LSDA}$= -13.5 \emph{GPa}. d, Calculated equations of state P(V) for LaMnPO in the tetragonal(blue) and orthorhombic (red)
structures. The dashed line corresponds to the computed critical pressure, -13.5 \emph{GPa}, where there is a coexistence between the tetragonal phase with V=154 \AA$^{3}$ to the orthorhombic phase that has V= 148 \AA$^{3}$.}
\end{figure}

How close is \mbox{LaMnPO} to an EDT?  Since this transition is often first order~\cite{mcwhan1973,limelette2003}, precursor effects may be minimal. To answer this question, we have determined the volume dependent electronic structure, with the expectation that the EDT will be evidenced by a collapse of the Mn moment at the critical volume. We use local spin density approximation (LSDA) calculations (Fig.~4a), as they are computationally less intensive than DFT+DMFT, yet they reproduce the qualitative features of the \mbox{LaMnPO} phase diagram, as shown below. The density of states (DOS) for the 3d Mn electrons in the ambient pressure tetragonal \mbox{LaMnPO} structure
shows an insulating behavior, although LSDA underestimates the gap, while the majority spin DOS lies almost completely below the Fermi level (Fig.~4a). This strong spin polarization is maintained at P=14.3 \emph{GPa},  although the gap has already closed. Accordingly, the Mn moment initially decreases slowly, but ultimately undergoes a more rapid collapse between 15-30 \emph{GPa} (Fig.~4b). We have carried out x-ray diffraction measurements on powdered \mbox{LaMnPO} at pressures as large as 43 \emph{GPa} to determine whether a volume collapse associated with the proposed EDT is present (Fig.~5a). We confirm that \mbox{LaMnPO} has the expected ZrCuSiAs structure at low pressures~\cite{nientiedt1997}, but new peaks are observed in the powder patterns above 16 \emph{GPa} (Fig.~5a). Structural refinements described in the Supplemental Information find that \mbox{LaMnPO} undergoes an orthorhombic distortion at 16 \emph{GPa} to the same structure type found in LaFeAsO at the lowest temperatures (insets Fig.~5c) ~\cite{mcguire2008}. This tetragonal-orthorhombic transition involves a sudden reduction of the \emph{c-axis} lattice constant by $\sim$4 $\%$, while the \emph{a} and \emph{b} lattice constants separate continuously with increasing pressure (Fig.~5b). A second pressure-driven transition is observed at 31 \emph{GPa}, which involves a sudden and large collapse of all three lattice constants that corresponds to a $\sim$ 12$\%$ reduction in the total volume, but with no apparent symmetry change.

\begin{figure}
\noindent\includegraphics[width=8.7cm]{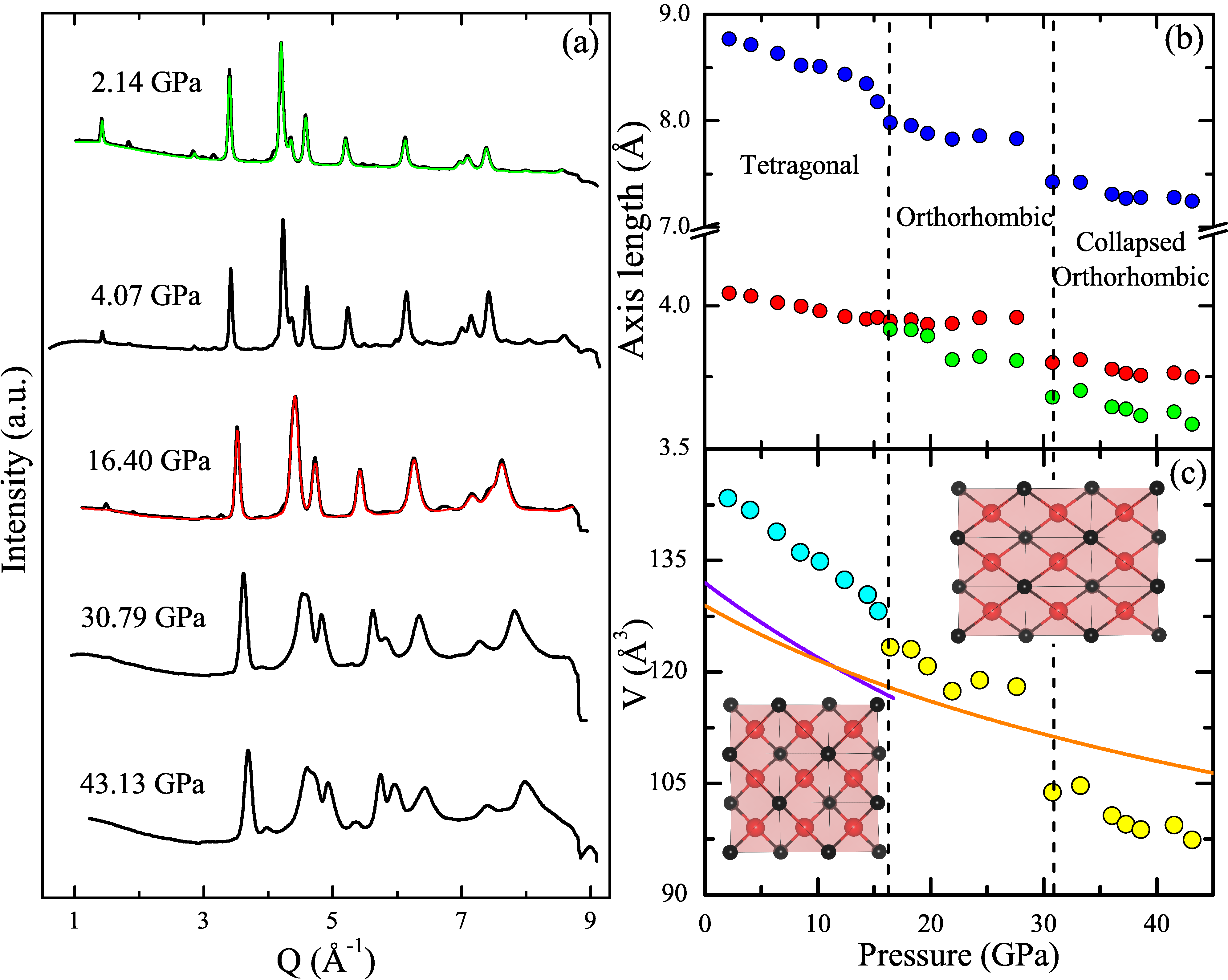}
\caption{The structure and lattice parameters of \mbox{LaMnPO} under pressure. a, An overlay of x-ray diffraction scans measured at several different pressures showing the appearance of new peaks at 16.40 \emph{GPa} and 30.79 \emph{GPa} .  A Rietveld refinement of the measurement performed at 2.14 \emph{GPa} is shown in green, and a Le Bail decomposition of the measurement performed at 16.40 \emph{GPa} is shown in red.  b, The a (red), b (green), and c (blue) lattice parameters that are taken from fitted peak positions are plotted as functions of pressure. c,  The experimentally determined cell volumes for the tetragonal (cyan) and orthorhombic (yellow) structures. The Murnaghan equation of state P(V) is computed from the cell volumes of the tetragonal structure (purple line) and the orthorhombic structure (orange line). The vertical dashed lines in b and c mark the structural transitions from tetragonal to orthorhombic at 16 \emph{GPa} and to the collapsed orthorhombic state at 31 \emph{GPa}.  The insetted images depict the crystal structure within the Mn-P plane as viewed along the c axis in both the tetragonal and orthorhombic states of LaMnPO with Mn drawn in red and P in black.  The orthorhombicity is exagerrated for clarity.\\*}
\end{figure}

The pressure dependencies of the enthalpies of the tetragonal and orthorhombic structures, derived from LSDA calculations of their respective total energies and equations of state (Fig.~4c),  cross at a pressure of -13.5 \emph{GPa}(Fig.~4d). The LSDA calculations and the equation of state P(V) (Fig.~4d) place the transition at a slightly expanded cell volume V$_{t-o}$=1.17 V$_{eq,LSDA}$=1.17(132\AA$^{3}$)= 154 \AA$^{3}$, although we note that LSDA finds P$_{eq}$ = -9 \emph{GPa} and not 1 bar for the tetragonal V$_{eq,LSDA}$. Accomodating this pressure offset suggests that the calculated critical pressure is more nearly  $\simeq$ -4.5 \emph{GPa}. While x-ray diffraction experiments find (Fig.~5c) that the transition actually occurs for a slightly compressed lattice with V$_{t-o}$= 130\AA$^{3}$=0.88 V$_{eq,expt}$=0.88 (147\AA$^{3}$), both theory and experiment agree that ambient pressure LaMnPO is near a tetragonal-orthorhombic transition. This transition has a major impact on the electronic and magnetic properties of LaMnPO. The orthorhombic structure computed with the cell volume appropriate for the experimental pressure P=16.4 \emph{GPa} is metallic, with both majority and minority spin DOS largely occupied (Fig.~4a), and accordingly the orthorhombic structure has a calculated moment that is smaller than that of the tetragonal structure, regardless of pressure (Fig.~4b). Consequently, the predicted tetragonal-orthorhombic transition is accompanied by a sharp reduction in moment at 16 \emph{GPa} from $\sim$ 2.7 $\mu_{B}$/Mn to 0.8 $\mu_{B}$/Mn. The computed moment continues to decrease as the orthorhombic structure is pressurized, until it finally vanishes near $\simeq$30 \emph{GPa}. It is tempting to conclude that the volume collapse observed in the X-ray diffraction near 31 \emph{GPa} is associated with this suppression of the Mn moment.

We have shown that DFT+DMFT provides an excellent account of the insulating gap and ordered moments that are measured in \mbox{LaMnPO}, while previous GGA+U calculations required different values of U to reproduce these results~\cite{yanagi2009}. These calculations as well as x-ray absorption measurements find substantial charge fluctuations, suggesting that LaMnPO is closer to an EDT than was previously appreciated. Electron doping scarcely affects the insulating gap, the Ne\'{e}l temperature, and the ordered moment, and no EDT is observed even for large doping levels, indicating that the charged and moment bearing in-gap states remain localized. However, pressure is effective at driving an EDT, and LSDA predicts that LaMnPO is very close to a tetragonal-orthorhombic transition that is accompanied by a sharp decrease in the Mn moment. X-ray diffraction measurements identify this structural transition at 16 \emph{GPa}, similar to the pressures required to make metallic LaFeAsO and BaFe$_{2}$As$_{2}$ superconducting~\cite{sefat2011}.  Surprisingly modest pressures are required to transform LaMnPO from a gapped and insulating tetragonal structure with a large Mn moment to a gapless and metallic orthorhombic structure with a small Mn moment that is comparable in magnitude to the Fe moments found in LaFeAsO~\cite{delacruz2008} and the AFe$_{2}$As$_{2}$ (A=Ca, Sr, Ba, Eu) compounds~\cite{huang2008}.  Pressure drives a transition in LaMnPO from the tetragonal to orthorhombic structures, which is the reverse of the low temperature transitions from orthorhombic to tetragonal or collapsed tetragonal structures that are found in pressurized LaFeAsO and the AFe$_{2}$As$_{2}$ (A=Ca, Sr, Ba, Eu) compounds~\cite{sefat2011}. Intriguingly, the net effect of pressure is much the same in LaMnPO and its Fe-based relatives, weakening the magnetic character of the ambient pressure compounds, which in the case of the Fe-based compounds is a necessary step towards producing superconductivity.



\begin{materials}
Single crystals of \mbox{LaMnPO} were synthesized from a NaCl-KCl eutectic flux, which produced very thin and platelike crystals with dimensions as large as 2-3 mm. Single crystal x-ray diffraction measurements confirmed that they form in the reported ZrCuSiAs structure~\cite{nientiedt1997} with the c-axis aligned perpendicular to the plate.  Pyrohydrolysis measurements were used to determine the absolute F concentrations for the series \mbox{LaMnPO}$_{1-x}$F$_{x}$~\cite{simonson2011}.  Electrical resistivity measurements were carried out in a Quantum Design Physical Property Measurement System, using a dc current of 100 nA flowing along the a-axis of a single crystal, as well as on a piece of the polycrystalline sample used for neutron diffraction.  Room temperature infrared (IR) transmission spectra were measured using a Bruker Vertex v/70 FT-IR spectrometer coupled to an IR microscope, which allowed us to obtain reliable data even on small ($<$ 1 \emph{mm}$^2$) crystals. A Quantum Design Magnetic Property Measurement System was used to perform magnetization measurements for T$\leq$400 \emph{K} on a collection of \mbox{LaMnPO} single crystals contained in a gold sachet, aligned with respect to their \emph{c-axes} but not within the \emph{ab} plane. The neutron diffraction sample was prepared using solid state synthesis, and powder x-ray diffraction found slight contamination by $\sim$2$\%$ La$_{2}$O$_{3}$, while magnetic susceptibility measurements found that $\sim$0.2$\%$ of ferromagnetic MnP was present. Neutron diffraction experiments used a high temperature furnace on the BT-9 triple axis spectrometer at the National Institute of Standards and Technology and a neutron wavelength of 2.36 \AA.  Angle-dispersive XRD experiments were carried out at the Beijing Synchrotron Radiation Facility (BSRF) on beamline 4W2 with a monochromatic x-ray beam of wavelength 0.6199 \AA, and the data were analyzed with Jana2006.  The photoemission and x-ray absorption experiments were performed at the Dragon beamline of the NSRRC in Taiwan, using an ultra-high vacuum system with a pressure in the low 10$^{-10}$ mbar range. The overall energy resolution in photoemission was set to 0.15 eV FWHM at 110 eV photon energy. The x-ray absorption spectra at the Mn $L_{2,3}$ edges were taken in the total electron yield mode with energy resolution of the photons of 0.3 eV. Before the measurements, the LaMnPO and MnO single crystals were cleaved \textit{in-situ} to obtain clean surfaces. The electronic structure of \mbox{LaMnPO} was determined using a combination of density functional theory and dynamical mean field theory (DFT+DMFT), which is based on the full-potential linear augmented plane wave method implemented in Wien2K. Identification of commercial equipment in the text is not intended to imply recommendation or endorsement by the National Institute of Standards and Technology.
\end{materials}


\begin{acknowledgments}
The authors are grateful for valuable discussions with J. W. Allen and  A. Nevidomskyy.
This research was supported by a Department of Defense National Security Science and Engineering Faculty Fellowship via the Air Force Office of Scientific Research. Part of this work was supported  by the National Science Foundation under Grant No. 1066293 and the hospitality of the Aspen Center for Physics. A.E. is supported by the European Union through the ITN SOPRANO network. L. Sun acknowledges the support of NSCF (10874230 and 11074294).
We acknowledge the support of the National Institute of Standards and Technology, U. S. Department of Commerce in providing the neutron research facilities used in this work.
\end{acknowledgments}





\end{article}








\end{document}